\documentclass[12pt]{article}
\usepackage[paperwidth=8.5in, paperheight=11in, top=1.5in, bottom=1.5in, margin=1in]{geometry}
\usepackage{tikz}
\usetikzlibrary{calc}
\usetikzlibrary{trees}

\usepackage{amsfonts}
\usepackage{amssymb}
\usepackage{amsmath}
\usepackage{amsthm}
\usepackage[multiple]{footmisc}

\usepackage{color}
\usepackage{fullpage} 
\usepackage{setspace}

\usepackage[titletoc,toc,title]{appendix}
\usepackage{titlesec}
\usepackage{natbib}

\usepackage[dvips]{epsfig}
\usepackage{epstopdf} 
\usepackage{amsthm}
\usepackage{pdfpages}

\usepackage{graphicx}
\usepackage{subcaption }

\usepackage[labelfont=bf]{caption}
\usepackage[font={footnotesize}]{caption}

\usepackage{paralist}
\usepackage{multirow}
\usepackage{rotating}

\newtheorem{theorem}{Theorem}

\newtheorem{proposition}[theorem]{Proposition}

\begin{document}

\title{\textbf{A Closed-Form Solution to the Risk-Taking Motivation of Subordinated Debtholders} \thanks{Yuval Heller: Bar-Ilan University, Ramat Gan, Israel, yuval.heller@biu.ac.il. Peleg Lazar: Tel Aviv University, Tel Aviv, Israel, sharonp5@post.tau.ac.il. Raviv: Bar-Ilan University, Ramat Gan, Israel, Alon.Raviv@biu.ac.il. Yuval Heller is grateful to the \textit{European Research Council} for its financial support (starting grant \#677057). Alon Raviv acknowledges the financial support of the Israel Science Foundation (ISF) through grant number 969/15.}} 
\bigskip
\author{Yuval Heller\\ Bar Ilan University
	\and
	Sharon Peleg Lazar\\ Tel Aviv University
	\and
	Alon Raviv\\ Bar Ilan University
}

\bigskip
\date{February 2019}
\maketitle

\bigskip
\begin{abstract}
	Black and Cox (1976) claim that the value of junior debt is increasing in asset risk when the firm's value is low. We show, using closed-form solution, that the junior debt's value is hump-shaped. This has interesting implications for the market-discipline role of banks' subdebt.
	
 
	\bigskip


\end{abstract}

\newpage	

\section{Introduction} \label{section: Introduction}

Over the past few decades the size and complexity of financial institutions has increased to an extent that challenges regulators' ability to monitor them. Many proposals suggest greater reliance on junior debt or subordinated debt (hereafter, subdebt) as a tool to discipline banks' risk-taking.\footnote{Another important tool for decreasing bank stock holders' risk taking motivation debated heavily in recent literature is contingent capital (CoCo). For examples of analysis of CoCo's effect on risk taking motivation see \citet{martynova2018convertible} and \citet{hilscher2014bank}.} It is argued that the negative effect of excessive risk-taking on subdebt price encourages subdebtholders to monitor banks closely in a way that is  aligned with the deposit insurer's incentive \citep{flannery2001faces}. Moreover, a change in the subdebt price can signal a change in the bank's risk, allowing the regulator and other market participants to discipline the bank \citep{evanoff2001sub, dewatripont1993efficient} 
 
The 2007--2009 financial crisis put into question the effectiveness of subdebt as a monitoring tool. \cite{calomiris2013design} claim that subdebt was ineffective in mitigating excessive risk due to  subdebtholders' lack of motivation to monitor banks' asset risk. Intuitively, when a bank is in financial distress, the only way its subdebtholders can be paid is by winning a large and risky bet; hence, the risk preferences of subdebtholders become more like those of shareholders. The risk preferences of the junior debt holders has been analyzed theoretically in \citet{black1976valuing} (and further developed in \citealp{gorton1990}), using the contingent claim approach for pricing corporate liabilities (\citealp{merton1974pricing}). They show that the value of a firm's junior debt can be expressed as a ``bull spread" position, which is composed of a long call option on the firm's asset with a strike price equal to the face value of the senior debt and a short call option with a strike  equal to the total debt of the firm. Specifically, \citet[p. 369]{black1976valuing} claim that when the firm's asset value is below a certain threshold (the discounted geometric mean of the face value of the deposits and the total debt), then the value of subdebt is an increasing function of asset  risk, since subdebt is then effectively the residual claimant (see also a similar claim in \citealp[p. 122]{gorton1990}).

In this note, we clarify and extend the above claim. We use a closed-form solution \citep[as in][]{black1976valuing} to show that when the firm's value is below the threshold mentioned above, the relationship between the subdebt value and the asset risk is hump-shaped (rather than increasing with asset risk), and to characterize  an interior level of risk that maximizes the value of subdebt.\footnote{We present a closed-form solution to the sensitivity of a bull spread option strategy to asset risk in \citet{peleg2017bank} and \citet{peleg2018risk}. However, the focus in that paper is on the sensitivity of the bank's stock  to asset risk when bank assets are a risky debt claim.} The revised analysis can have important implications for the expected effect of subdebt on the risk-taking of financial institutions in time of distress. Specifically, the subdebtholders will not be motivated to increase risk as much as possible (which is the stockholders's motivation); rather, they will be motivated to increase risk to some intermediate level. Thus, if the subdebtholders can affect risk-shifting, then they will choose a moderate level of risk and risk-shifting will be limited to the level that maximizes the value of the subdebt.\footnote{The question whether and to what extent subdebtholders can affect the choice of asset risk is a separate question that is beyond the scope of this paper. In \citet{heller2019banks} we apply a  game-theoretic bargaining analysis to study the equilibrium risk induced by joint control of the bank by shareholders and subdebtholders.} Moreover, we show that the level of asset risk that maximizes the value of the subdebt is an increasing function of the firm's leverage and of the proportion of the senior debt out of the firm's total debt. 

Our analysis can explain the empirical results in \citet{ashcraft2008does}, which documents that an increase in the amount of subdebt in regulatory capital has a positive effect in helping a bank recover from financial distress. Others also document that including subdebt in a bank's capital reduced risk-shifting during the crisis period \citep{nguyen2013disciplinary, john2010outside, belkhir2013subordinated,chen2011subordinated}.

\section{Analysis} \label{section: Model}

We use the expanded contingent claims valuation model derived by \citet{black1976valuing} to encompass the case of multiple debt claims (junior and senior). Our model and notations closely follow \citet{gorton1990}.

\subsection{Valuation} \label{subsection: valuation} 

We begin by describing the firm's liability structure and expressing the values of the different claims. 

A firm is funded by stock with market value $S$, and two types of debt with different priorities as claimants, such that at debt maturity the junior debt is repaid only after the senior debt is repaid in full. Both loans are zero-coupon loans maturing at time $T$. The senior loan's face value is $F_S$ and its market value is $B_S$ and the junior debt's face value is $F_J$ and its market value is $B_J$. The value of the firm's assets, $V$, follows a geometric Brownian motion under the risk-neutral measure according to the following equation: $dV_{t}= r \cdot V_{t} \cdot dt +\sigma \cdot V_{t} \cdot dW$, where $r$ is the instantaneous risk-free rate of return, $\sigma$ is the instantaneous volatility of asset value, and $dW$ is a standard Wiener process under the risk-neutral probability measure. 
 
If at debt maturity the value of the firm's assets is greater than the sum of the face values of all its debt, $F_S+F_J$, then both debtholders are repaid in full and equityholders receive the residual. By contrast, if the firm's asset value at maturity is below the sum of the face values of all its debt, but above the face value of the senior debt, $F_S< V < F_S+F_J$, then the senior debt is repaid in full and the junior debtholders are the residual claimants (equityholders do not receive any payoff). Otherwise, if the firm's asset value is below the face value of the senior debt, $V<F_S$, the senior debtholders receive all the asset value and the junior debtholders and equityholdesr are not repaid at all.

Hence, the senior debtholders' payoff at maturity is the minimum between the firm's asset value and the face value of their debt: $B_{S,T} = \min \{V_T, F_S\}$. This expression can be rearranged and expressed as $B_{S,T} = F_S-\max \{F_S-V_T, 0\}$. As discussed in \citet{merton1974pricing}, this payoff is equivalent to the payoff of a risk-free debt with face value $F_S$ and a short position in a European put option. Therefore, the present value of the senior debt is
\[ B_{S,t} =  F_S \cdot e^{-r (T-t)} -Put_t(V_t,F_S,\sigma, T-t,r), \]
where $Put_t(V_t,F_S,\sigma, T-t,r)$ is the value of a European put option on the firm's asset value at time $t$, with a strike price equal to the face value of senior debt $F_S$, asset risk $\sigma$, and time to maturity $T-t$. Under the above described geometric Brownian motion, the value of the option  can be found using the \citet{black1973pricing} equation.

The junior debt's payoff at maturity is the minimum between the value of assets left after repaying the senior debt, if any, and the face value of the junior debt $\min \{V_T-F_S, F_J\}$, as long as the payoff is nonnegative, i.e., $B_{J,T} = \max\{\min \{V_T-F_S, F_J\},0\}$. This payoff can be rearranged and expressed as $B_{J,T} = \max \{V_T-F_S,0\} -\max \{V_T-(F_S + F_J),0 \}$, which is equivalent to a long position in a European call option with a strike price equal to the face value of the senior debt, $F_S$, and a short position in a European call option with a strike price equal to the sum of the face values of all of the firm's debt, $F_S+F_J$. Therefore, the value of the junior debt prior to maturity is
\begin{equation}
B_{J,t} = Call_t(V_t, F_S, \sigma, T-t, r) - Call_t(V_t, F_S+F_J, \sigma, T-t, r),\label{eqution: value_junior}
\end{equation}
where $Call_t(V_t, \cdot, \sigma, T-t, r)$ is the value of a European call option according to the \citet{black1973pricing} equation.

The stock's payoff at debt maturity is $S_T = \max \{V_T-(F_S+F_J), 0\}$. This payoff  can be replicated by a European call option on the value of the firm's assets, with a strike price equal to the sum of the face values of all the firm's debt \citep{galai1976option}. Therefore, the value of stock at time $t$ is
\[ S_t = Call_t(V_t,F_S+F_J, \sigma, T-t, r).
\label{equation: value_equity}\]
The value of the firm's assets and the payoff to each of its claimholders at debt maturity is presented in Figure \ref{fig: payoffs}.

\subsection{Junior Debtholder's Asset Risk Preferences} \label{subsection: preferences}


As pointed out by \citet[p. 122]{gorton1990}, ``If the promised payment of the senior debt is close to the value of the firm, then junior debt is effectively the residual claimant and will behave like an equity claim. If, however, the value of the firm is significantly higher than the promised payment on the senior debt, then the junior debt will behave like debt." More precisely, the sensitivity of the value of junior debt to the level of asset risk is divided into two segments defined by the threshold (\citealp[][Eq. 10]{black1976valuing}; \citealp[][Eq. 7]{gorton1990}):
\begin{equation}
	\hat{V}\equiv e^{- \left(r + \frac{\sigma^2}{2}\right)(T-t)} \sqrt{ F_S \cdot \left(F_S+F_J \right) }, \label{eq: V_hat} 
\end{equation}
which is a function of the geometric mean of the face value of senior debt and the sum of the face values of all of the firm's debt. \citet[][p. 360]{black1976valuing} conclude that ``Analysis of the function shows $J$ [the bond's value] is an increasing (decreasing) function of $\sigma^2$ for $V$ less than (greater than) $\hat{V}$."  

We clarify this statement by noting that when the value of the borrower's assets is below this threshold, the relationship between the market value of the subdebt claim and asset risk is \textit{hump-shaped}, and the maximum value of the subdebt claim is reached at the level of asset risk defined as follows (the proof is presented in the Appendix):
\begin{equation}
	\sigma^{max} \equiv \arg\max_{\sigma} B_{J,t} (\sigma)=
	\sqrt{\frac{1}{T-t} \ln\left( \frac{F_S \cdot (F_S+F_J)}  {V_t^2}\right) -2r },
	\label{eq: sigma_max_sub}
\end{equation} 
if $V\leq  V^* \equiv e^{-r \cdot (T-t)} \sqrt{F_S \cdot (F_S+F_J)}$; otherwise there is no internal solution and $\sigma^{max}=0$. The level of asset risk that maximizes the value of the junior debt, $\sigma^{max}$, is an increasing function of the firm's leverage ($F_S+F_J$ to $V$) and the ratio of senior debt to asset value ($F_S$ to $V$). 

The following result summarizes the above analysis of the value of the bank stock as a function of the asset's risk.\footnote{ Under the assumption of continuous dividends, $qV_t dt$, we find that the both the threshold under which risk shifting takes place and the optimal level of risk when risk shifting takes place are slightly higher and equal $\hat{V}\equiv e^{- \left(r -q + \frac{\sigma^2}{2}\right)(T-t)} \sqrt{ F_S \cdot \left(F_S+F_J \right) }$ and $\sigma^{max} \equiv\sqrt{\frac{1}{T-t} \ln\left( \frac{F_S \cdot (F_S+F_J)}  {V_t^2}\right) -2r +2q}$.}
\begin{proposition} \label {proposition: sigma_max_subdebt}
	The value of the junior debt is (1) decreasing with asset risk if $V_t>V^*$ and (2) hump-shaped (unimodal) in asset risk if $V_t<V^*$, and, in this case, its maximum is obtained for risk level $\sigma^{max}$. Moreover, the risk level that maximizes the value of junior debt is higher than the initial risk (i.e., $\sigma^{max}>\sigma_{0}$) if and only if  $V_t<\hat{V}$. 
\end{proposition}

\begin{figure}
	\centering	
	\begin{subfigure}[]{0.7\textwidth}
		\includegraphics[width=\textwidth]{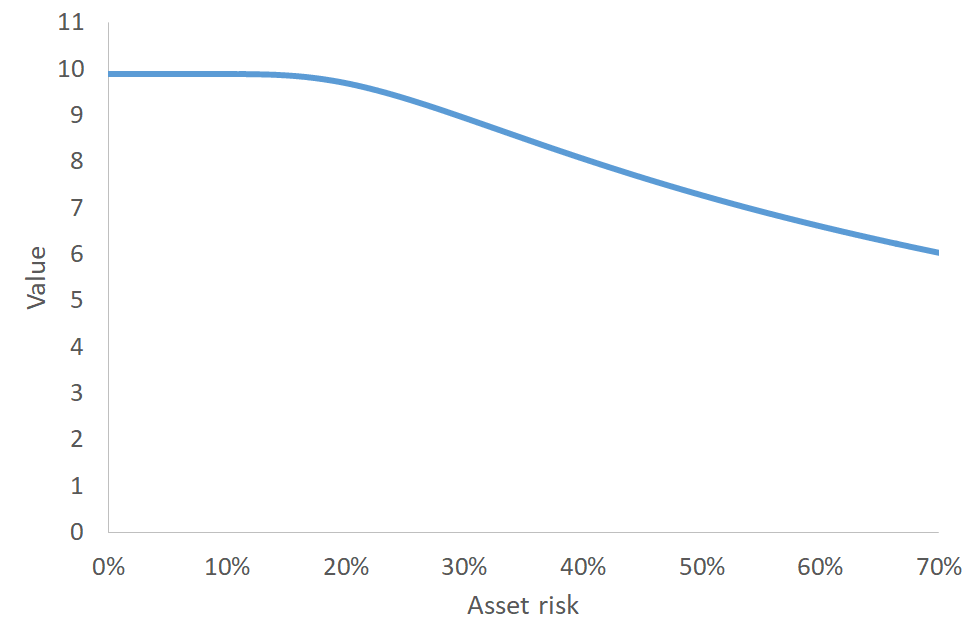}
		\caption{Asset value: $V=100$} \label{fig:value_function_of_risk100}
	\end{subfigure}
	\hfill
	\begin{subfigure}[]{0.7\textwidth}
		\includegraphics[width=\textwidth]{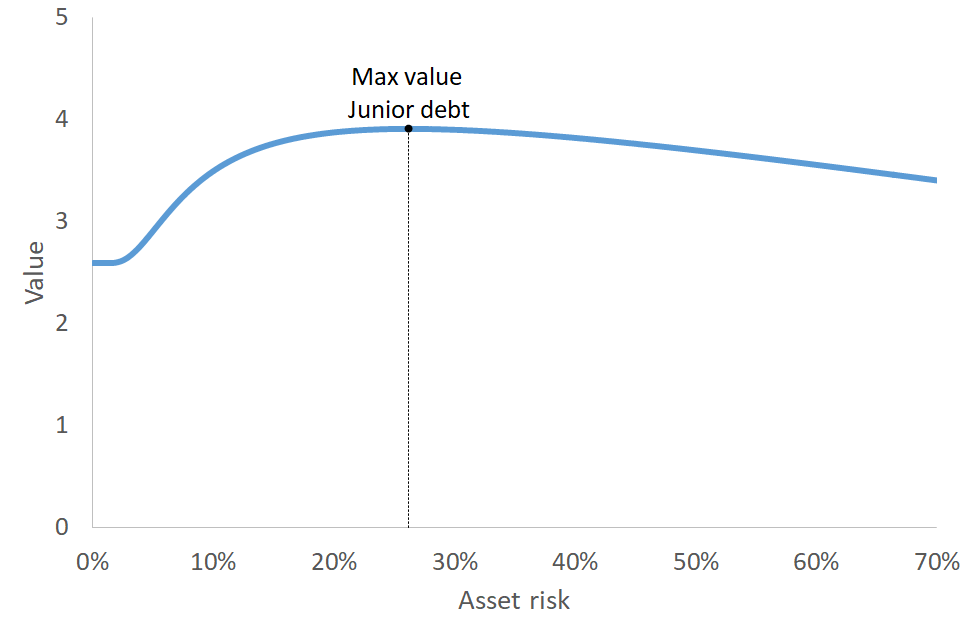}
		\caption{Asset value: $V=62$} \label{fig:value_function_of_risk62}
	\end{subfigure}
	\caption{\textbf{The value of junior debt as a function of the level of asset risk.} The face value of the senior debt is $F_S=60$ and the face value of the junior debt is $F_J=10$. In addition, the time to maturity is one year and the risk-free rate is $r=1\%$. Given these values, $\hat{V}=63.8$.} 
	\label{fig:value_function_of_risk}
\end{figure}

The proposition is demonstrated in Figure \ref{fig:value_function_of_risk}. In panel (\subref{fig:value_function_of_risk100}) the value of assets is above both $\hat{V}$ and $V^*$ and the value of junior debt is decreasing with asset risk. By contrast, in panel (\subref{fig:value_function_of_risk62}) the value of junior debt is hump-shaped with respect to asset risk and achieves its maximum value when asset risk equals $\sigma^{max}= 26.2\%$.

\begin{figure}[]
	\centering	
	\includegraphics[width=0.8\textwidth]{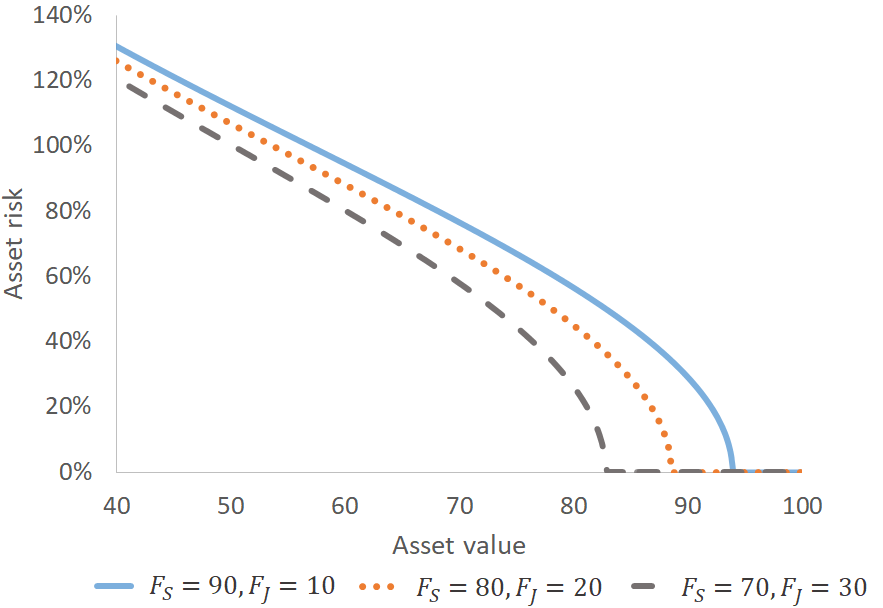}
	\smallskip
	\caption{\textbf{The level of asset risk that maximizes the value of junior debt for different proportions of junior debt.} The figure depicts the equilibrium level of asset risk chosen by the junior debtholders. The different lines (ranging from 10\% to 30\%) correspond to different proportions of junior debt out of the total debt. The sum of the face values of both the senior debt and the junior debt is fixed at $F_S+F_J=100$. The initial level of asset risk is $\sigma_0=10\%$. In addition, the time to maturity is one year and the risk-free rate is $r=1\%$.}
	\label{fig:leverage}
\end{figure}

\paragraph{The effect of the proportion of junior debt} 
We find that both the range of asset values where risk-shifting takes place and the level of asset risk that maximizes the value of junior debt decrease with the proportion of junior debt.
To see this, note that both $\hat{V}$ (Eq. \ref{eq: V_hat}) and $\sigma^{max}$ (Eq. \ref{eq: sigma_max_sub}) decrease when the face value of junior debt $F_J$ increases while the sum of the face values of the total debt $F_S+F_J$ remains unchanged (where an increase in $F_J$ while keeping $F_S+F_J$ unchanged implies a decrease in $F_S$).
This result is demonstrated in Figure \ref{fig:leverage}, where the sum of the face values of the total debt is constant and equal to $F_S+F_J=100$ while the face value of the junior debt changes. 

\newpage
\bibliographystyle{econometrica}
\bibliography{subordinated_debt}

\newpage
\titleformat{\section}{\large\bfseries}{\appendixname~\thesection :}{0.5em}{}
\appendix
\section{Proofs} \label{appendix}
\numberwithin{equation}{section}
\setcounter{equation}{0}

The value of junior debt (Eq. \ref{eqution: value_junior}) is equivalent to a portfolio of two call options on the value of the firm's assets
$ B_{J,t} = Call_t(V_t, F_S, \sigma, T-t, r) - Call_t(V_t, F_S+F_J, \sigma, T-t, r). $

To find the level of asset risk that maximizes the value of junior debt we calculate the derivative of the value of junior debt with respect to asset risk. Since it is well known that
\[\frac{\partial Call_t}{\partial \sigma}= \frac{\sqrt{T}}{\sqrt{2\pi}} \cdot V \cdot e^{-\frac{1}{2} \cdot (d_1(F_i))^2},\]
where 
$d_1(f_i) = \frac{1}{\sigma\sqrt{T-t}} \cdot \left[\ln\left(\frac{V_t}{F_i}\right) + \left(r + \frac{1}{2} \sigma^2 \right)\cdot(T-t)\right]$,
we get
\[\frac{\partial  B_{J,t}}{\partial \sigma}= \frac{\sqrt{T}}{\sqrt{2\pi}} \cdot V \cdot e^{-\frac{1}{2} \cdot (d_1(F_S))^2} - \frac{\sqrt{T}}{\sqrt{2\pi}} \cdot V \cdot e^{-\frac{1}{2} \cdot (d_1(F_S+F_J))^2}. \]

After rearranging it can be shown that the derivative equals
\[\frac{\partial  B_{J,t}}{\partial \sigma}= \frac{\sqrt{T-t}}{\sqrt{2\pi}} \cdot V_t \cdot  e^{-\frac{1}{2 \cdot T \cdot \sigma^2}}
\left[e^a-e^b \right], \]

where $a$ and $b$ are defined as
\[ a= -2\cdot\ln{V_t} \cdot \ln{F_S}+(\ln({F_S}))^2 - 2\cdot\ln\left(F_S\right) \cdot \left(r + \frac{\sigma^2}{2}\right) \cdot (T-t) \]

\[b= -2\cdot\ln{V_t} \cdot \ln{F_S+F_J}+(\ln({F_S+F_J}))^2 - 2\cdot\ln\left(F_S+F_J\right) \cdot \left(r + \frac{\sigma^2}{2}\right) \cdot (T-t).\]

The payoff is maximized with respect to asset risk in cases where the first derivative equals zero. This happens when either $V_{C,t}=0$ or $a=b$. Since the first option is of no interest economically we focus on the second option. We find that $a=b$ when 
\begin{equation}
V_t = e^{-\left(r + \frac{1}{2}\sigma^2\right)\cdot (T-t)} \cdot  \sqrt{F_S \cdot (F_S+F_J)},  
\label{eq:Vstardtar}
\end{equation}
Based on Equation \ref{eq:Vstardtar}, we define $\hat{V}$ as the borrower's asset value for which the claim's value is maximized given a level of asset risk $\sigma$. This threshold is identical to the one found in \citet[][Eq. 10]{black1976valuing} and \citet[][Eq. 7]{gorton1990}. We also note that the derivative changes its sign from positive to negative above the threshold, $\hat{V}$, meaning that the bank's claimholder would like to increase risk below that level and to decrease it above that level of assets.

Next, fixing the level of assets, we find that $a=b$ when
\begin{equation}
\sigma =
\sqrt{\frac{1}{T-t} \ln\left( \frac{F_S \cdot (F_S+F_J)}{\left(V_t\right)^2}\right) -2r }. \label{eq:sigmamax}
\end{equation}

However, for both Equations \ref{eq:Vstardtar} and \ref{eq:sigmamax} to hold, that is, for an internal solution to exist, the firm's asset value must be below $V^*$ defined as
$V^* \equiv e^{-r \cdot (T-t)} \sqrt{F_S \cdot (F_S+F_J)} $.

\end{document}